\documentstyle[aps,prd,twocolumn]{revtex}

\input epsf

\begin{document}

\title{Constant Curvature Black Holes}

\author{M\'aximo Ba\~nados }

\address{Centro de Estudios Cient\'{\i}ficos de Santiago, Casilla 16443,
Santiago, Chile, and\\ Departamento de F\'{\i}sica, Universidad de
Santiago de Chile, Casilla 307, Santiago 2, Chile}

\maketitle

\begin{abstract}
Constant curvature black holes are constructed by identifying points in
anti-de Sitter space. In $n$ dimensions, the resulting topology is
$\Re^{n-1} \times S_1$, as opposed to the usual $\Re^2 \times S_{n-2}$
Schwarzschild black hole, and the corresponding causal structure is
displayed by a $(n-1)-$dimensional picture, as opposed to the usual
2-dimensional Kruskal diagram. The five dimensional case, which can be
embedded in a Chern-Simons supergravity theory, is analyzed in
detail.  
\end{abstract}

The family of black objects has grown considerably during the last decade.
Black holes, black strings and black branes have been studied in various
dimensions and theories, and it is now commonly believed that they play an
important role in string theory - the most promising theory to unify
gravity with the other fundamental forces.   

Black holes, black branes and black strings have event horizons, a
null surface in spacetime beyond which light cannot escape. This
property is best displayed in terms of the causal structure. The
standard Kruskal picture (without charge or angular momentum) is drawn in
Fig. 1. The curved lines represent the future and past singularities and
the diagonals the horizon. A future-directed observer in region II will
necessarily hit the singularity since in order to go back to region I
which is connected to infinity, he or she would need a velocity greater
than light. In the simplest situation,  each point in Fig. 1 represents a
$(n-2)-$sphere with $n$ the dimension of spacetime. The spacetime topology
is thus $\Re^2\times S_{n-2}$. 


\begin{center}
\leavevmode
\epsfysize=3.5cm

\hspace{.6cm}\epsfbox{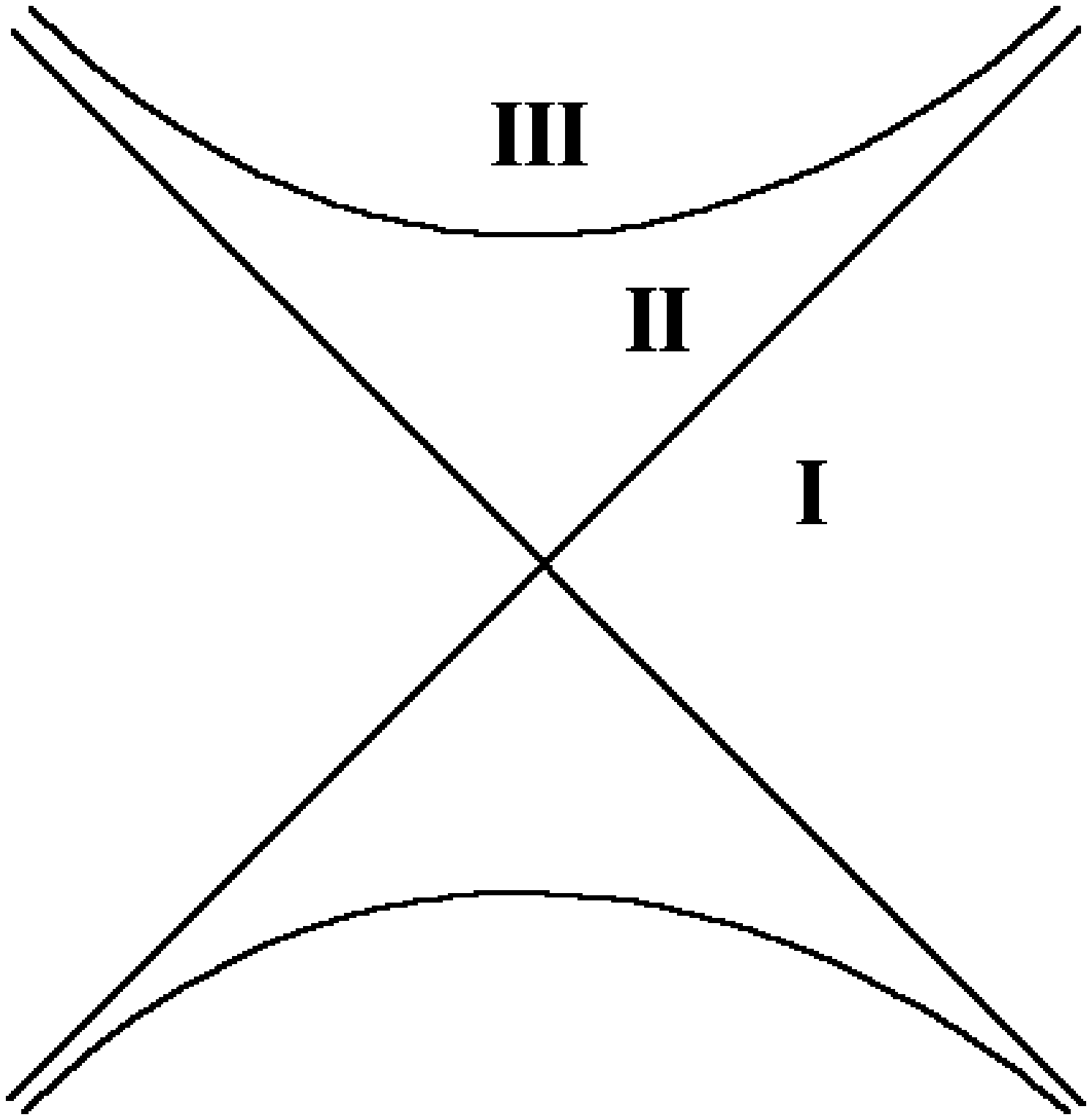} 

{\bf Figure 1.} 
 \end{center}  

\begin{quote}
{\footnotesize
Standard Kruskal diagram. Region I is connected to infinity and region II
is the interior of the black hole.}
\end{quote}


In this note we shall construct metrics with constant negative curvature
whose Kruskal diagram will be given by the picture shown in Fig. 2. The
hyperboloid represents the singularity, while the cone represents the
horizon.  Drawing the two surfaces together one obtains a picture
equivalent to Fig. 1 rotated around the z-axis. An observer falling into
the region in between the hyperboloid and the cone cannot escape back
because, as before, that would require a velocity greater than light. In
that sense, Fig. 2 displays a black hole. 

In our construction, 
each point in Fig. 2 will represent a circle and the spacetime topology is
thus $\Re^3\times S_1$. In an arbitrary dimension $n$, Fig. 2 will be
replaced by its natural $n-1$ generalization and the black hole will then
have the  topology $\Re^{n-1} \times S_1$.

  \begin{center}
 \leavevmode
 \epsfysize=3.5cm

\hspace{.3cm}\epsfbox{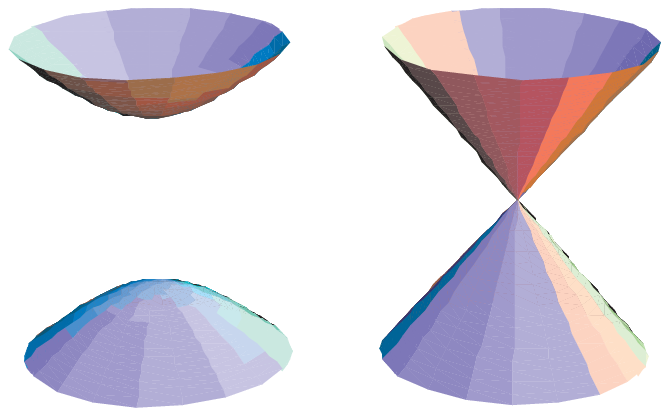}

 {\bf Figure 2}.
\end{center}
\begin{quote}
{\footnotesize 
The hyperboloid represents the singularity and the hypercone the horizon.
Drawing the two surfaces together gives Fig. 1 rotated around the z-axis.}  
\end{quote}


The existence of a ``$\Re^{n-1} \times S_1$ black hole" was first
suggested in \cite{BTZu}, and also extensively discussed with M. Henneaux.
The main two problems faced on that occasion, which are addressed here,
were an apparent change of signature when crossing the horizon and the
definition of conserved charges associated to asymptotic symmetries.

The construction of the $\Re^{n-1} \times S_1$ black hole is a natural
extension of the procedure yielding the 2+1 black hole from anti-de Sitter
space with identified points\cite{O,exbh}. We shall then start by
briefly reviewing that construction \cite{BHTZ}, in the non-rotating case.

In three dimensions, anti-de Sitter space is defined as the universal
covering of the surface,
\begin{equation}
-x_0^2 + x_1^2 +x_2^2 - x_3^2 = -l^2.
\label{ads/3}
\end{equation}    
This surface has six Killing vectors, 2 rotations and 4 boosts. Pick
up the Killing vector $\xi =(r_+/l)( x_2 \partial_3 + x_3 \partial_2)$
whose
norm  $\xi^2 = (r_+^2/l^2)(-x_2^2 + x_3^2)$ can be negative, zero or
positive ($r_+$ is an arbitrary real number).  The surface
(\ref{ads/3}) can be rewritten as 
\begin{equation}
x_0^2 = x_1^2 + l^2 (1 - \xi^2/r_+^2 ),
\label{bh3}
\end{equation} 
and can be plotted parametrically in terms of the values of
$\xi^2$. One finds that the semiplane $x_0|x_1$, $x^0+x^1>0$, has 3
regions separated by $\xi^2=0$ and $\xi^2=r_+^2$. Indeed, the picture is
the same as in Fig. 1 on which $\xi^2=0$ represents the singularity and
$\xi^2=r_+^2$ the horizon:
\begin{eqnarray}
\mbox{I}\ \ &:&\ \  r_+^2 < \xi^2 <\infty, \nonumber \\  
\mbox{II}\ \ &:&\ \   0< \xi^2 < r_+^2, \nonumber \\
\mbox{III}\ \ &:&\ \   -\infty <\xi^2 \leq 0. \nonumber
\end{eqnarray}
Also, each point in Fig. 1 represents a one dimensional non-compact
manifold: the values of $x_2$ and $x_3$ with $x_2^2-x_3^2$ (the norm of
the Killing vector) fixed. (Regions I,II and III are repeated in the lower
semiplane $x^0+x^1<0$.)

This, of course, does not transform anti-de Sitter space into a black
hole. What does produce the black hole is the identification of
points along the orbit of $\xi$. Since $\xi$ is a Killing vector one
can produce a new metric with constant curvature by taking the
quotient of the surface (\ref{ads/3}) with a discrete subgroup
generated by $\xi$. If we do so, region III ($\xi^2 \leq 0$) acquires a
pathological cronological structure and therefore it must be cut off
from the physical spacetime\cite{BHTZ}. In that sense, the surface
$\xi^2=0$ becomes a singularity. Moreover, the non-compact one dimensional
manifold orthogonal to Fig. 1 becomes isomorphic to $S_1$. The quotient
space has thus the topology $\Re^2\times S_1$ and the induced metric is
the 2+1 black hole\cite{BHTZ}.

The above picture has a natural generalization to higher
dimensions \cite{O,exbh}. In $n$ dimensions anti-de Sitter space is
defined as (the universal covering) of  the surface
\begin{equation}
-x_0^2 + x^2_1 + \cdots + x^2_{n-2} + x_{n-1}^2 - x_n^2 = -l^2.
\label{ads/n}
\end{equation}
Consider the boost $\xi =(r_+/l) (x_{n-1} \partial_{n} + x_n
\partial_{n-1})$ with
norm $\xi^2=(r_+^2/l^2)(-x_{n-1}^2+x_n^2)$. As before, we plot
parametrically the surface (\ref{ads/n}) in terms of the values of
$\xi^2$. The horizon $(\xi^2=r_+^2)$ is now represented by the hypercone,
\begin{equation}
x_0^2=x_1^2 + \cdots + x_{n-2}^2,
\label{h/n}
\end{equation}
while the singularity $(\xi^2=0)$ is represented by the hyperboloid,
\begin{equation}
x_0^2=x_1^2 + \cdots + x_{n-2}^2 + l^2,
\label{s/n}
\end{equation}
just as illustrated in Fig. 2.  We now identify points along the orbit of
$\xi$ obtaining the desired causal structure.  The region behind the
hyperboloid ($\xi^2<0$) has to be removed from the physical spacetime
because it contains closed timelike curves. The hyperboloid is thus a
singularity because timelike geodesics end there.  Again, each point in
Fig. 2 represents a circle (the identified line) and the topology of the
quotient space is $\Re^{n-1}\times S_1$.  

To go further in the discussion, let us introduce local  
coordinates on anti-de Sitter space (in the region $\xi^2>0$) adapted to
the Killing vector used to make the identifications. We introduce the
$n$ dimensionless local coordinates $(y_\alpha,\phi)$ by, 
\begin{eqnarray}
x_\alpha &=& \frac{2l y_\alpha}{1-y^2}, \mbox{\hspace{.6cm} }
\alpha=0,...,n-2
\label{y} \\
x_{n-1}  &=& \frac{lr}{r_+}  \sinh\left(\frac{r_+\phi}{l}\right),
\nonumber \\ 
    x_n  &=& \frac{lr}{r_+}  \cosh\left(\frac{r_+\phi}{l}\right),
    \nonumber 
\end{eqnarray}
with 
\begin{equation}
r = r_+ \frac{1+y^2}{1-y^2}
\end{equation}
and $y^2 = \eta_{\alpha\beta}\, y^\alpha y^\beta $
[$\eta_{\alpha\beta}=\mbox{diag}(-1,1,...,1)$].  The
coordinate ranges are
$\infty < \phi < \infty$ and $-\infty < y^\alpha <\infty$ with the
restriction $-1<y^2<1$.

The induced metric has the Kruskal form,
\begin{equation} 
ds^2 =  \frac{l^2(r+r_+)^2}{r_+^2}\, dy^\alpha
dy^\beta\eta_{\alpha\beta} + r^2 d\phi^2, 
\label{ds/krus}
\end{equation}
and the Killing vector reads $\xi =\partial_\phi$ with
$\xi^2=r^2$. The
quotient space is thus simply obtained by identifying $\phi \sim \phi+2\pi
n$, and the resulting topology is $\Re^{n-1}\times S_1$. With the help of
(\ref{y}), it is clear that the Kruskal diagram associated to this
geometry is the one shown in Fig. 2 extended to an arbitrary dimension
$n$. Thus, the metric (\ref{ds/krus}) represents the $\Re^{n-1}\times S_1$
black hole written in Kruskal coordinates.  
Note also that the above metric is a natural generalization of the
2+1 black hole. Indeed, setting $n=3$ in (\ref{ds/krus}) gives the 
non-rotating 2+1 black hole metric written in Kruskal coordinates
\cite{BHTZ}.

Hereafter we shall restrict the discussion to the five dimensional
case which has some special features that will be explained below. 
One may wonder if there exists Schwarzschild coordinates for the above
metric. We shall see that they exists only in the exterior region. In
particular one cannot find Schwarzschild coordinates interpolating the
inner and outer regions.  

Let us introduce local ``spherical" coordinates
($t,\theta,\chi,r$) in the hyperplane $y^\alpha$:
\begin{eqnarray}
y_0 = f \cos{\theta} \sinh{(r_+t/l)}, &\mbox{\hspace{.6cm} }& 
           y_2 = f \sin{\theta} \sin{\chi},   \nonumber\\    
y_1 = f \cos{\theta} \cosh{(r_+t/l)}, &\mbox{\hspace{.6cm} }&
        y_3 = f \sin{\theta} \cos{\chi} \label{cc},
                      \nonumber
\end{eqnarray}
with $f(r) = [(r-r_+)/(r+r_+)]^{1/2}$.  [Note that these coordinates, with
ranges $0 < \theta<\pi$, $0\leq\chi<2\pi$, $-\infty<t<\infty$ and
$r_+<r<\infty$, do not cover the whole region $r>r_+$ but only 
$-1<y_2,y_3<1$.]  The metric (\ref{ds/krus}) adquires the Schwarzschild
form,  
\begin{equation}
ds^2 = l^2 N^2 d\Omega_3 +  N^{-2} dr^2 + r^2 d\phi^2, 
\label{ds/sch}
\end{equation}
with $N^2(r) = (r^2 -r_+^2)/l^2 $ and
\begin{equation}
d\Omega_3 = - \cos^2{\theta}\, dt^2 + \frac{l^2}{r_+^2} (d\theta^2 +
\sin^2{\theta} d\chi^2) 
\label{sph}
\end{equation}
The horizon in these coordinates is located at $r=r_+$, the point
where $N^2$ vanishes.  It should now be clear why
the Kruskal diagram is 4 dimensional (in five dimensions): the ``lapse" 
function $N^2$ multiplies not a single coordinate but the whole
(hyperbolic) three sphere $d\Omega_3$. It is also clear that one cannot
find Schwarzschild coordinates interpolating the outer and inner regions.
For $r<r_+$, the metric (\ref{ds/sch}) changes its signature and
therefore it does not represent the interior of the black hole.   

The black hole just constructed has an Euclidean sector which can be
obtained by setting $\tau = it$ in (\ref{ds/sch}), or $y_0 \rightarrow
iy_0$ in (\ref{ds/krus}). In this sector the coordinates
$(t,\theta,\chi,r)$ do cover the whole Euclidean black hole spacetime
which, as usual, is isomorphic to the exterior of the Minkowskian black
hole. 

For later use we point out that the boundary of the spatial sections has
the topology $S_2\times S_1$.  This is easily seen by setting
$t,r=const.$ in (\ref{ds/sch}), or, alternatively, by setting
$y_0,r=const.$ in (\ref{ds/krus}).  

Just as in 2+1 dimensions, angular momentum in the plane $t|\phi$ can be
added by considering a different Killing vector to do the identifications. 
Here, we shall not carry the complete geometrical construction which for
dimensions greater than three is not trivial. Instead,
we shall add a new charge by boosting the above metric in the plane
$t|\phi$.  This is most easily done by setting $r_+=l$ in (\ref{ds/sch}),
making the replacements,
\begin{eqnarray}
t   &\rightarrow& \beta t\frac{r_+}{l^2} +  (\phi - \Omega \, \beta t)
\frac{r_-}{l},
\\
\phi &\rightarrow& \beta t\frac{r_-}{l^2} + (\phi - \Omega \,
\beta t) \frac{r_+}{l},
\label{ang}
\end{eqnarray}
($r_+>r_-$ arbitrary constants), and identifying points along the new
angular coordinate $\phi$: $\phi \sim \phi + 2\pi n$. The parameters
$\beta$ and $\Omega$ are introduced here only to make the canonical
structure of the global charges explicit: $\beta$ is the
conjugate to the energy while $\Omega$ in the conjugate to the
angular momentum.  One could set $\beta=l$ and $\Omega=0$ without altering
the physical properties of the solution at all. 

The explicit form of the resulting metric is not very illuminating so we
do not include it here.  We only point out that the constant $r_+$
parameterizes the location of the outer horizon, and that the new metric
has two independent conserved charges (see below). In the Euclidean
formalism, the time coordinate $\tau=-it$ must be periodic in order to
avoid conical singularities. This gives the value $\beta = (2\pi
r_+l^2)/(r_+^2-r_-^2)$ [with $0\leq t <1$] which can be interpreted as the
inverse temperature of the black hole.

Since the above geometries are locally anti-de Sitter, they are natural
solutions of Einstein equations with a negative cosmological constant.
However, due to the non-standard asymptotic behavior of (\ref{ds/sch}) one
finds that all conserved charges are infinite\cite{Marc}.  This is a
serious obstacle since if no physical conserved charges can be defined the
physical relevance of the solution is not clear. 

We would like to point out here that global conserved charges associated
to these black holes can be defined in the context of a Chern-Simons
supergravity theory in five dimensions purposed sometime ago by
Chamseddine \cite{Chamseddine}.  This action is constructed as a
Chern-Simons theory for the supergroup $SU(2,2|N)$ \cite{Chamseddine} and
it represents a natural extension of the three dimensional supergravity
theory constructed in \cite{Achucarro}. The explicit form of the
action is not needed here (it can be found in \cite{Chamseddine}). 
What is more useful are the equations of motion which read
(setting all fermions and non-Abelian gauge fields equal to zero),
\begin{eqnarray}
l\epsilon_{abcde} \tilde R^{ab} \wedge \tilde R^{cd} - 2F\wedge T_e 
                 &=& 0\label{e},\\
2\epsilon_{abcde} \tilde R^{ab} \wedge T^c + l F\wedge \tilde R_{de}  
                 &=& 0,
\label{w}\\
\tilde R^{ab} \wedge \tilde R_{ab} -2l^{-2} T^a\wedge T_a  + \Delta\
F \wedge F &=& 0,
\label{b}
\end{eqnarray}
where $\Delta =  N^{-4} - 4^{-4}$ and $N$ is the number of fermions in the
action. $T^a=D e^a$ is the 2-form torsion, and 
\begin{equation}
\tilde R^{ab} = R^{ab} + l^{-2} e^a \wedge e^b
\end{equation}
where $R^{ab}=dw^{ab}+w^a_{\ c} \wedge w^{cb} $ is the 2-form Lorentz
curvature. Finally, the 2-form $F=dA$ is the curvature associated to the
$U(1)$ gauge field $A$ which is necessary to achieve supersymmetry. The
above equations of motion are not equivalent to the five dimensional
Einstein equations. However, note that in the sector $F=T^a=0$ and small
curvatures, $R^2\approx 0$, they do reduce to Einstein equations.  

Note that the group $SU(2,2|N)$ has as a subgroup $SO(4,2) \times U(1)$.
In fact, the above equations can be interpreted as a Chern-Simons theory
for the group $SO(4,2)\times U(1)$.  Also, since $SO(4,2)$ is isomorphic
to the anti-de Sitter group in five dimensions (whose Lie algebra is
generated by $J_{ab}$ and $P_a$), $\tilde R^{ab}$ and $T^a$ are,
respectively, the projections of the $SO(4,2)$ curvature along $J_{ab}$
and $P_a$. 

The dynamics of Chern-Simons theories has been studied in general in
\cite{BGH2}. In particular, it was shown in \cite{BGH2} that those
theories with a gauge group $G\times U(1)$ enjoy a drastic simplification
in the canonical and asymptotic structure if: (i) the last term in
(\ref{b}) is identically zero and, (ii) $F$ has maximum rank
\cite{BGH2}. Condition (i) is a restriction on the invariant tensor
necessary to construct the Chern-Simons theory and (ii) ensures the
existence of local degrees of freedom. We shall then consider here the
theory with $N=4$ ($\Delta=0$) and study solutions for which $F$ has
rank 4 (in five dimensions, the maximum rank of a 2-form is 4). 

It can now be seen that the geometries described at the beginning
of this note solve the above equations of motion (for $N=4$) because they
have constant curvature ($\tilde R^{ab} =0$) and zero torsion ($T^a=0$).
Note also that $F$ is left arbitrary and therefore it can have
maximum rank.    

Global conserved charges associated to the above equations are easily
constructed \cite{BGH2}. Let $E= E^a P_a + (1/2) E^{ab} J_{ab}$ be the
left hand side of (\ref{e}) and (\ref{w}), and let $\delta e^a,\delta
w^{ab}$ be perturbations of the vielbein and spin connection. It follows
that $\delta E$ is a covariantly conserved current of the classical
history. Now, let $\lambda=\lambda^a P_a + (1/2)\lambda^{ab} J_{ab}$ be a
Killing vector of the background configuration ($D \lambda=0$ where $D$ is
the anti-de Sitter covariant derivative), then $\delta J=Tr(\lambda \delta
E)$ is a conserved current in the ordinary sense. As in any gauge theory,
$\delta J$ is a total derivative and integrating it 
over a spatial section $\Sigma$ with boundary $\partial \Sigma$ one
obtains, 
\begin{equation}
\delta Q[\eta_a,\eta_{ab}] = \frac{1}{8\pi^2}\int_{\partial \Sigma} F
\wedge (2 \eta_a \delta e^a - l\eta_{ab} \delta w^{ab}).
\label{Q}
\end{equation}
This formula holds under the boundary condition 
$\tilde R^{ab} =0=T^a$ which, of course, is satisfied by our
solution. The normalization factor $8\pi^2$ comes from the volume of
$\partial\Sigma =S_2\times S_1$. 

The formula (\ref{Q}) depends on the 2-form $F$ which was not determined
by the equations of motion. The only local conditions over $F$ are $dF=0$
(since $F=dA$), and that it must have maximum rank \cite{BGH2}. In
particular, $F$ must be different from zero everywhere.  It turns out that
global considerations suggest a natural choice for the pull back of $F$
into $\partial \Sigma=S_2\times S_1$, namely, proportional to the area
2-form of $S_2$, $F = k\, \sin\theta d\theta \wedge d\chi$. Note that
$dF=0$ implies that $k$ is constant over $S_1$.

The formula for the charge thus becomes,
\begin{equation}
\delta Q[\eta_a,\eta_{ab}] = \frac{k}{8\pi^2} \int_{\partial\Sigma}
( 2 \eta_a \delta e^a_\phi - l\eta_{ab} \delta
w^{ab}_\phi)dS  
\label{Q2}
\end{equation}
where $e^a_\phi$ and $w^{ab}_\phi$ are the projections of the veilbein
and spin connection along $S_1$ (parameterized by $\phi$), and
$dS=\sin\theta d\theta d\chi d\phi$. 

The black hole geometries described above have two commuting Killing
vectors, $\partial_t$ and $\partial_\phi$, whose conserved charges can be
associated, respectively, to the energy and angular momentum of the
solution. Due to the flatness of the anti-de Sitter curvature in our
solution, a diffeomorphism with parameter $\xi^\mu$ is equivalent to
an anti-de Sitter gauge transformation with parameter $\xi^\mu [e^a_\mu
P_a+(1/2) w^{ab}_\mu J_{ab}]$ \cite{Witten88}. Hence, the Killing
displacement $\partial_t$ can be replaced by a gauge transformation with
parameter $e^a_t P_a + (1/2) w^{ab}_t J_{ab}$, and $\partial_\phi$ by a
gauge transformation with parameter $e^a_\phi P_a + (1/2) w^{ab}_\phi
J_{ab}$. 

Computing the energy $M$ and angular momentum $J$ as described
above one finds,
\begin{eqnarray}
M = \frac{1}{l} Q[e^a_t,w^{ab}_t]       &=& k \frac{2r_+ r_-}{l^2},                     
                   \label{M}\\
J = Q[e^a_\phi,w^{ab}_\phi] &=& k\frac{r_+^2 + r_-^2}{l},
\label{J}
\end{eqnarray}
where we have normalized the time Killing vector by $\beta =l$, and
$\Omega=0$ (no shift at infinity).  In obtaining $M$ and $J$ we have use
the boundary condition $\delta k=0$ \cite{BGH2}.  Note that $k$, which
acts
as a coupling constant, is not an universal parameter in the action but
the -fixed- value of the $U(1)$ field strength $F$. This is similar to
what happens in String Theory where the coupling constant is equal to the
value of the dilaton field at infinity.   

Comparing the above values for $M$ and $J$ with those obtained in the 2+1
theory \cite{BHTZ}, one discovers that they are reversed. One can easily
trace back the reason for this interchange between mass and angular
momentum: in the action, the only term that contributes to the conserved
charges is $A \wedge (l^2\tilde R^{ab} \wedge \tilde R_{ab} - 2T^a \wedge
T_a)$ which is not the usual Hilbert Lagrangian. This term, which involves
the $SO(4,2)$ Pontryagin density as opposed to the Euler density, is not
parity invariant unless $A$ is a density. Note also that the black hole
horizon exists only for $J \geq M$, as opposed to the standard bound
$M\geq J$ \cite{CT}. The permutation of charges is a well known phenomena
when a duality transformation is applied.  In this case, the ``duality
transformation" comes from the fact that the relevant part of the action
was constructed with $\eta_{ab}$-tensors instead of the 5D Levi-Cevita
symbol $\epsilon_{abcde}$. It is worth mentioning here that the same
situation is observed in 2+1 dimensions if one considers, instead of the
usual Einstein-Hilbert action, the ``exotic" 2+1 action \cite{Witten88}
which is also constructed using $\eta_{ab}$ tensors instead of the Levi
Civita symbol. Another example where this phenomena occurs was reported in
\cite{Carlip-Gegemberg-Mann}.

The black holes constructed here have mass and angular momentum. It
is now natural to compute their entropy. In our case, the quickest way to
arrive at the right result is by computing the entropy as a Euclidean
Noether charge at the horizon\cite{Wald}. Imposing at the horizon $\beta =
(2\pi l^2 r_+) /(r_+^2 - r_-^2)$ and $\Omega=r_-/(lr_+)$, ensuring the
absence of conical singularities, one finds,
\begin{equation}
S = Q[e^a_t(r_+),w^{ab}_t(r_+)]=  4\pi k\, r_-.
\label{S}
\end{equation}      
This result is rather surprising because it does not give an entropy
proportional to the area of $S_1$ ($2\pi r_+$).  This is not a
contradiction. In fact, for Lagrangians with higher order terms in
the curvature the entropy is not proportional to the area
\cite{Jacobson}. In this case, besides the higher order curvature
terms, the relevant term in the action has a non-standard parity
with respect to the geometric variables.  [In the model discussed in
\cite{Carlip-Gegemberg-Mann} the same value for the entropy was found.]
The entropy given in (\ref{S}) satisfies the first law, 
\begin{equation}
\delta M = T \delta S + \Omega \delta J, 
\label{fl}
\end{equation}
where $M$ and $J$ are given in (\ref{M},\ref{J}) and $T=1/\beta$. 

During this work I have benefited from many discussions and useful
comments raised by Andy Gomberoff, Marc Henneaux, Claudio Teitelboim and
Jorge Zanelli. This work was partially supported by the grant \# 1960065
from FONDECYT (Chile), and institutional support by a group of Chilean
companies (EMPRESAS CMPC, CGE, COPEC, CODELCO, MINERA LA ESCONDIDA,
NOVAGAS, ENERSIS, BUSINESS DESIGN ASS. and XEROX Chile).

\end{document}